\documentclass[twocolumn,preprintnumbers]{revtex4}
\usepackage{graphicx,caption,float}
\begin{document}

\title{Non-Ergodic Complexity Management}
\author{Nicola Piccinini $^{1}$, David Lambert  $^{1}$, Bruce J. West  $^{2}$, Mauro Bologna  $^{3}$, Paolo Grigolini  $^{1}$}

\affiliation{1) Center for Nonlinear Science, University of North Texas, P.O. Box 311427, Denton, Texas 76203-1427\\
2) Information Science Directorate, Army Research Office, Research Triangle Park, North Carolina, 27709, USA\\
3) Instituto de Alta Investigation, Universidad de Tarapac\'{a}, Casilla 6-D, Arica, Chile}

\begin{abstract}
Linear response theory (LRT), the backbone of non-equilibrium statistical
physics, has recently been extended to explain how and why non-ergodic
renewal processes are insensitive to simple perturbations \cite{aquino10},
such as in habituation. It was established that a permanent correlation
resulted between an external stimulus and the response of a complex system
generating non-ergodic renewal processes, when the stimulus is a similar
non-ergodic process. This is the principle of complexity management (PCM),
whose proof relies on ensemble distribution functions \cite{aquino11}.
Herein we extend the proof to the non-ergodic case using time averages and a
single time series, hence making it usable in real life situations where
ensemble averages cannot be performed because of the very nature of the
complex systems being studied.
\end{abstract}

\maketitle
   
The mathematician Norbert Wiener, in the middle of the last century \cite%
{wiener48}, speculated that a system high in energy can be controlled by one
that is low in energy. The necessary force is produced by the low energy
system being high in information content, and the high energy system being
low in information content. Consequently, there is an information gradient
that produces the force by which the low energy system controls the high
energy system, through a flow of information against the traditional energy
gradient. Quantifying the transfer of information from a complex system high
in information to one low in information is the first articulation of a
universal principle of network science and we refer to this speculation as
Wiener's Rule.

In a modern context Wiener's Rule can be understood as an entropic force,
used to explain such diverse phenomena as the elasticity of freely-jointed
polymer molecules \cite{neumann77}, oceanic forces \cite{holloway09} and the
conscious states in the human brain, through neuroimaging \cite{harris14}.
Over the past decade the nascent field of network science has been applied
to determining the conditions under which the Wiener's Rule is facilitated
or suppressed. After half a century Wiener's Rule has been shown to be
correct and has been superseded by the more detailed Principle of Complexity
Management \cite{aquino10,aquino11}.

One result of the many analyses of information transfer, that is being
continually rediscovered, is that complex networks in living systems exist
at, or on the edge of, phase transitions (collective, cooperative behavior),
which optimizes both intra- and inter-network information transmission \cite%
{beggs03}. Moreover, the statistical distributions of a diverse collection
of complex systems are inverse power law, whether modeling the connectivity
of the internet or social groups, the frequency or magnitude of earthquakes,
the number of solar flares, the time intervals in conversational turn
taking, and many other phenomena, see for example \cite{west11}. The
power-law index is the measure of complexity in each system.

Traditional methods of non-equilibrium statistical physics have not been
successful in addressing the question of information transfer between
complex networks. For example, in studying the response of complex systems
to harmonic perturbations it was determined by many authors, among which are 
\cite{sokolov01,barbi05,weron08,magdziarz08}, that \textit{linear response
theory} (LRT), a cornerstone of physics, was "dead". An assessment of this
premature death was made by Aquino \textit{et al}. \cite{aquino10,aquino11},
resulting in a generalization of LRT (GLRT) that was successfully applied to
the question of information transfer. In their discussion these latter
authors focused on the intimate connection between neural organization and
information theory, as well as the production of $1/f$ noise. This gave
solid ground to the observation that 1/f signals are encoded and transmitted
by sensory neurons with greater efficiency than are white noise signals \cite%
{yu05}. Psychologists interpret the generation of $1/f$ noise as a
manifestation of cognition \cite{gilden01,orden05}, although no
psychologically well founded model for the origin of $1/f$ noise yet exists 
\cite{farrell06}. However, experimental observation of brain dynamics either
monitoring EEG activity \cite{allegrini15} or through actigraphy \cite%
{ochab14} confirm that the awake condition of the brain is a source of $1/f$
noise \cite{allegrini09}.

Despite its successes, GLRT suffers from a fundamental limitation that hinders its application to many real world systems.
In this letter we review the current results obtained using GLRT, overcome
its fundamental limitation using theoretical arguments and verify the theory using numerical
simulations.

It is useful to introduce the notion of a \emph{renewal event, }which is an
event associated with a reorganization of the system under study. It is
customary to call the time between two renewal events a \emph{laminar region}.
As the word ``renewal" suggests, the lengths of two consecutive laminar
regions are independent. Our study includes the many complex systems that
exhibit dynamical behavior described by inverse power-law statistical
distributions. A good approximation for the waiting time distribution (WTD)
between two renewal events in these systems (equivalent to the distribution
of the lengths of the laminar regions) is : 
\begin{equation}
\psi (t)=\frac{(\mu -1)T^{\mu -1}}{(T+t)^{\mu }}  \label{eqWtd}
\end{equation}%
where $T$ and $\mu $ are parameters characterizing the complex system under
study. The power-law index $\mu $ is also called the index of complexity; it
must be larger than one in order for $\psi$ to be normalizable. With simple
calculations it can be shown that the second moment of the WTD is finite
when $\mu >3$. Thus, systems with $\mu$ in this range satisfy the central
limit theorem, hence they are in the Gauss basin of attraction \cite{annunziato01}.
When $2<\mu<3$ the second moment is infinite, so these 
systems obey the generalized central limit theorem \cite{annunziato01}
and are in the L\'{e}vy basin of attraction. Finally, when $1<\mu<2$,
the mean time also becomes infinite; in this case the generalized
central limit theorem does not apply. These systems are \emph{non-ergodic} as will be
made clear subsequently.

Equation (\ref{eqWtd}) can be used to calculate the probability of having a
laminar region that is at least as long as $t$ (survival probability): 
\begin{equation}
\Psi (t)=1-\int_{0}^{t}\!\psi (t)\,\text{\textrm{d}}t=\left( \frac{T}{T+t}\right) ^{\mu -1}  \label{eqSp}
\end{equation}

Another useful quantity that will play a key role in this Letter is the rate $R(t)$
at which new events are generated, given that an event occured at $t=0$.
When $2<\mu <3$ we have \cite{allegrini15b}
$R(t)\approx \overline{t}^{-1}[1+T^{\mu -2}(3-\mu )^{-1}t^{2-\mu }]$, where $\overline{t}$ is the
first moment. In this case the system is Poissonian only in the infinite
time limit. Finally, in the non-ergodic regime, Feller \cite{feller71}
demonstrated that the rate at which new events are generated is: 
\begin{equation}
R(t)\propto t^{\mu -2},\quad 1<\mu<2  \label{fellercascade}
\end{equation}
In this case the system is often referred to as \emph{non-Poissonian}. The
main implication of this result is that a system with $1<\mu <2$ is in a
perennial non-equilibrium state, as the rate at which events are generated
keeps decreasing forever (notice the difference with the usual Poissonian
case, where this rate is constant). A direct consequence of Eq.(\ref%
{fellercascade}) is that performing ensemble averages of statistical
properties, related to renewal events for systems that have an event at $t=0,
$ is different from making time averages of the same properties on a single system
that was prepared at $t=0$, since the latter averages change with time.
This change of statistical properties with time is a consequence of the fact that they are linked to the rate
of event generation. In other words, by definition, these latter systems are non-ergodic, as we
anticipated while discussing the properties of the moments of $\psi$.

In order to create a time series $\xi (t)$ for a complex system
characterized by the above statistical properties, a value 1 or -1 is
associated with each laminar region. At each renewal event a fair coin is
tossed to decide wether to switch from one value to the other. The time
series $\xi (t)$ allows us to define the autocorrelation function $\Phi
(t,t^{\prime })\equiv \left\langle \xi (t)\xi (t^{\prime })\right\rangle $
that is needed when the LRT and GLRT are introduced.

As an aside, we notice that the power spectrum of $\xi (t)$ also depends on $%
\mu $. In the Gauss basin of attraction \cite{annunziato01}, $\mu >3$, the
spectrum $S(f)$ for $f\ll 1$ is very flat as $\mu \rightarrow \infty $. For $%
2<\mu (=3-\beta )<3$, in the asymptotic region $t\gg T$, we have $%
S(f)\propto 1/f^{\beta }$ \cite{annunziato01}, which is $1/f$ noise with $\beta <1$. When $\mu <2
$, we have \cite{lukovic08} $S(f)\propto L^{\mu -2}f^{-\beta }$, where $L$
is the length of the time series. We also notice that, under the conditions $%
t\ll T/(\mu -2)$ and $\mu >2$, we have $S(f)\propto 1/f^{2}$, the same
result as that obtained for flicker noise.

As we stated before, a complex system characterized by the properties
described above does not respond to a periodic perturbation, hence the idea
that LRT was mistakenly believed to be ``dead". Aquino \textit{et al}. \cite%
{aquino10,aquino11} demonstrated that LRT can be generalized and
successfully applied GLRT to the case of one complex system perturbing
another. In the following, the former is denoted by P
(perturbing system), while the latter is denoted by S (responding system).
Thus, the S-system is characterized by the global variable $\xi _{S}(t)$ and
is perturbed by the global variable $\xi _{P}(t)$. Conventional LRT \cite%
{kubo85} is given by: 
\begin{equation}
\left\langle \xi _{S}(t)\right\rangle =\epsilon \int_{0}^{t}\chi
(t,t^{\prime })\xi _{P}\left( t^{\prime }\right) dt^{\prime }  \label{LRT}
\end{equation}

where the symbol $\left\langle \xi _{S}(t)\right\rangle $ denotes the Gibbs
ensemble average over infinitely many realizations of the response of $\xi
_{S}(t)$ to $\xi _{P}(t)$. Without loss of generality, in the absence of
perturbation this average is assumed to vanish. $\epsilon<<1$ is the
stimulus strength. LRT predicts the response of S on the basis of the
unperturbed autocorrelation function $\Phi _{S}(t,t^{\prime })$ of $\xi
_{S}(t)$. In fact, the function $\chi (t,t^{\prime })$, called the linear
response function (LRF), is related to the derivative of the autocorrelation
function, normalized so that its quadratic mean value is one. In LRT the
autocorrelation function is assumed to depend only on the difference between 
$t^{\prime }$ and $t$ (hence it is stationary, by definition), consequently
the derivative with respect to either time, $t$ or $t^{\prime }$, can be
taken, differing only by a change of sign \cite{kubo85}.

However, when the statistics are non-stationary the generalized LRF (GLRF)
is \cite{aquino10}: 
\begin{equation}
\chi \left( t,t^{\prime }\right) =\frac{d\Phi _{S}(t,t^{\prime })}{%
dt^{\prime }}=R_{S}\left( t^{\prime }\right) \Psi _{S}(t-t^{\prime }).
\label{LRF}
\end{equation}%
Where the subscript indicates that the rate of generation of new events $R(t)
$, the autocorrelation function $\Phi (t)$ and the survival probability $%
\Psi (t)$, are those of the resonding system. The Principle of Complexity
Management (PCM) is obtained by studying the cross correlation between $\xi
_{S}$ and $\xi _{P}$, normalized to $\epsilon $, as a function of $\mu _{S}$
and $\mu _{P}$, as $t\rightarrow \infty $: $\Phi _{\infty
}=\lim\limits_{t\rightarrow \infty }\left\langle \xi _{S}(t)\xi
_{P}(t)\right\rangle /\epsilon $. The calculations made by Aquino \textit{et
al}. \cite{aquino10,aquino11} show a number of remarkable properties. For
example, if the S-system is ergodic and the P-system is non-ergodic, the
cross-correlation is maximum: this means that there is a flux of information
from the P-system to the S-system (Wiener's Rule). When the P-system is
ergodic and the S-system is non-ergodic, the asymptotic cross-correlation
vanishes; thus, there is no residual response of the S-system to the
P-stimulus. Note that this was the domain that earlier investgators
prematurely interpreted as the death of LRT. In the case in which both
systems are ergodic, there is a partial positive correlation between S and P
that changes with $\mu _{S}$ and $\mu _{P}$; as is the case when both
systems are non-ergodic.

The extraordinary results obtained using the asymptotic cross-correlation
function have a fundamental limitation, however, because the predictions of
this form of PCM rely on ensemble averages. Thus,
the predictions based on the cross-correlation are not necessarily valid when we have only a single
non-ergodic time series for each system, that is, when we cannot apply the equivalence between ensemble
averages and time averages. This is a common situation, since many interesting systems cannot
be replicated. Consider the response of a single molecule to its
environment \cite{singlemolecule} or a single brain to a unique stimulus: in both examples the
response time series is one of a kind.

\begin{figure*}[t]
\centering
\includegraphics[width= 180mm]{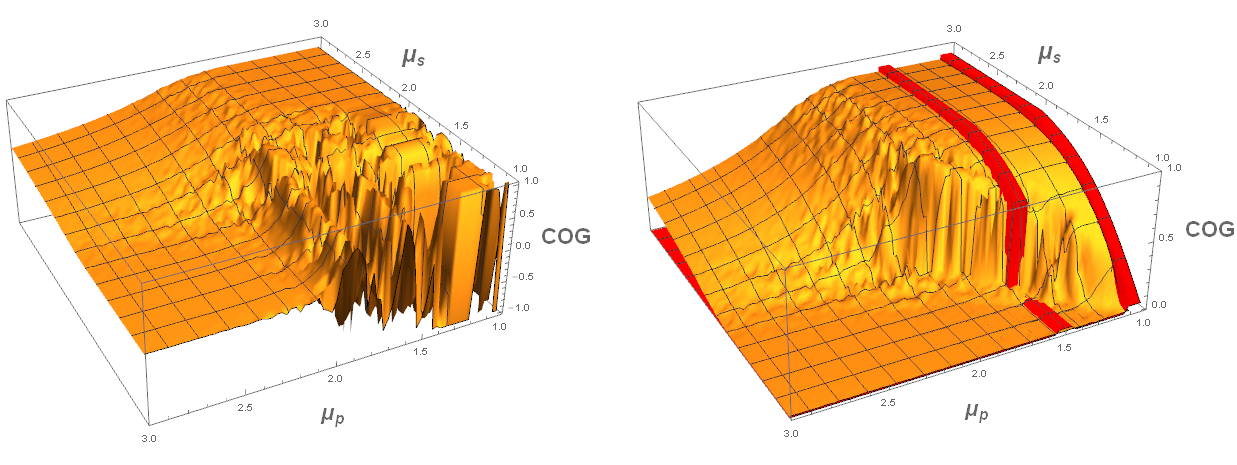}
\caption{(Color online) COG as a function of the inverse power-law indexes $\protect\mu_S$ and $\protect\mu_P$. Left panel: The time sequence of length $L$ is divided into $L/T_{W}$ intervals of length $T_W$. The
cross-correlation function of Eq. (\ref{integral}) is evaluated for each interval
and the landscape is obtained plotting the mean of the resulting
distribution of values. Right panel: The times $t_{i}$ of Eq. (\ref{integral}) are
the times of event occurrence and the landscape is obtained by plotting the
mean of the resulting distribution of values. The exact prediction of Eqs. (\ref{exact}) and (\ref{eqProbS})
is shown by the two red lines.}
\label{fig_twocubes}
\end{figure*}

We begin to address the limitation of a single time series by describing how the
S-system is stimulated. Recalling Eq. (\ref{eqWtd}), we note that there are
two parameters that can be perturbed: $\mu $ and $T$. Since $\mu $
quantifies the complexity of the system, it is reasonable to expect that it
can be forced to change only in response to very strong stimulation. A
non-invasive perturbation, therefore, is expected to only change $T$. This
restriction is in keeping with the dynamical approach to LRT used in \cite%
{silvestri09}, to design the GLRT \cite{aquino10,aquino11} that led to such
remarkably good agreement with experimental observation.

The P-system exerts its influence on the S-system as follows: if S has an
event at time $t$ and if its next laminar region is assigned a value with
the same sign as $\xi _{P}(t)$, then S is perturbed so that its
next laminar region tends to be longer, by assigning to its parameter $T$ in
Eq. (\ref{eqWtd}) the value $T_{+}=T(1+\epsilon )$. On the contrary, if the
next laminar region of S has a value with the opposite sign to that of $\xi
_{P}(t)$, then the value $T_{-}=T(1-\epsilon )$ is used, thus tending to
make the next laminar region shorter.

In order to assess the influence of P on S for a single time series, using
this perturbation procedure, it is natural to consider a time window of size 
$T_{W}$ and analyze the time averaged cross-correlation function: 
\begin{equation}
C\left( t_{0},T_{W}\right) \equiv \frac{1}{T_{W}}%
\int_{t_{0}}^{t_{0}+T_{W}}dt^{\prime }\xi _{S}(t^{\prime })\xi
_{P}(t^{\prime }).  \label{integral}
\end{equation}

By moving the starting point $t_{0}$ of the window and evaluating $C$,  a
density plot for the time averaged cross-correlation can be
created as a function of the power-law indices.
A measure of the influence of the P-system on the S-system is the center of
gravity (COG) of this density plot. In the domain $1<\mu _{S},\mu _{P}<2$,
the COG of the density plot is erratic; in sharp contrast with the smooth
behavior found in the calculations of the cross-correlation function in this
region obtained using ensemble distribution functions by Aquino et al. \cite%
{aquino10,aquino11}. This is clearly shown in the left panel of Fig. (\ref%
{fig_twocubes}), where $C/\epsilon $ is plotted as a function of $\mu _{S}$
and $\mu _{P}$. It is worth noting that different realizations of the figure
lead to different landscapes in the non-ergodic quadrant. The reasons behind
this behavior will become clear shortly.

The main contribution of this Letter has two parts. The first part is a new
data processing prescription that enables one to eliminate the erratic
behavior observed in the left panel of Fig. (\ref{fig_twocubes}) and produce the
smooth behavior of the right panel. In the second part we provide a theoretical justification for this prescription and
calculate the asymptotic cross-correlation function analytically.

The prescription is to locate the beginning $t_{0}$ of the window at which each $C$ is evaluated on
an event of either the perturbing or the perturbed system.

\begin{figure*}[t]
\includegraphics[width= 90mm]{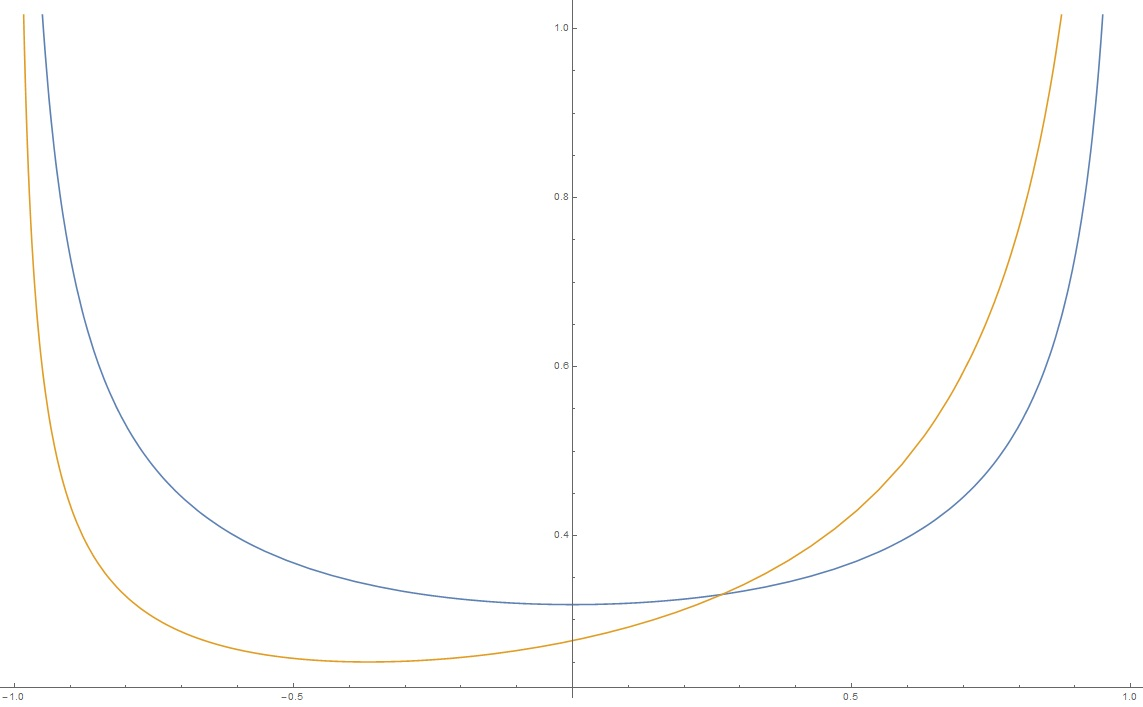}
\caption{(Color online) Unperturbed ($\protect\epsilon =0$) and perturbed ($%
\protect\epsilon =0.5$) Lamperti distributions with $\protect\mu =1.5$}
\label{figLamperti}
\end{figure*}

We now turn our attention to presenting the theoretical foundations that led
to the data processing prescription given above.
We start by considering the following random quantity:
$\overline{\xi _{S}}=\int_{0}^{t}dt^{\prime }\xi _{S}(t^{\prime })/t$ based on
different realizations of the unperturbed $\xi _{S}$ that was prepared so as
to have an event at $t_{0}=0$. Notice that the beginning of the window is
always located at $t=0$, in contrast to Eq.(\ref{integral}). In the case of $%
\mu _{S}<2$, it was shown by performing ensemble averages \cite%
{annunziato01} that $\overline{\xi _{S}}$ is characterized by the
Lamperti probability density function \cite{lamperti58}: 
\begin{equation}
\Pi (\overline{\xi _{S}})=\frac{2}{\pi }\frac{\left( 1-\overline{\xi _{S}}%
^{2}\right) ^{\alpha -1}sin\pi \alpha }{\left( 1-\overline{\xi _{S}}\right)
^{2\alpha }+\left( 1+\overline{\xi _{S}}\right) ^{2\alpha }+\left( 1-%
\overline{\xi _{S}}^{2}\right) ^{\alpha }cos\pi \alpha },  \label{ML2}
\end{equation}%
where $\alpha $ is $\mu -1$. Whose graph is depicted as the symmetric curve
in Fig. (\ref{figLamperti}). We notice that this distribution is clearly
non-ergodic as a single realization is most probably located around 1 or -1,
while the ensemble average is zero. 

We now consider the time-averaged quantity $\overline{\xi }\equiv \frac{1}{%
T_{W}}\int_{0}^{T_{W}}dt\xi _{S}(t)\xi _{P}(t),$ that is Eq. (\ref{integral}%
) when $t_{0}=0$ and we employ the same procedure followed in the
calculation of $\overline{\xi _{S}}$. Bologna \textit{et al} \cite{bologna10}%
, as well as Akimoto \cite{akimoto12}, demonstrated that the resulting
distribution is a skewed Lamperti distribution given by: 
\begin{equation}
\Pi (\overline{\xi })=\frac{2}{\pi }\frac{\left( 1-\overline{\xi }%
^{2}\right) ^{\alpha -1}sin\pi \alpha }{\left( 1-\overline{\xi }\right)
^{2\alpha }\eta +\left( 1+\overline{\xi }\right) ^{2\alpha }\frac{1}{\eta }%
+\left( 1-\overline{\xi }^{2}\right) ^{\alpha }cos\pi \alpha }.  \label{U}
\end{equation}

The parameter $\eta $ is responsible for the asymmetry of the curve in Fig. (%
\ref{figLamperti}) and is related to the intensity of the perturbation $%
\epsilon $ by: 
\begin{equation}
\eta \equiv \left( \frac{1-\epsilon }{1+\epsilon }\right) ^{\alpha }.
\label{parameter}
\end{equation}%
The COG of the density plot is given by 
\begin{equation}
B\left( \mu _{P},\mu _{S}\right) =\epsilon \frac{1-\left( \frac{1-\epsilon }{%
1+\epsilon }\right) ^{\mu _{S}-1}}{1+\left( \frac{1-\epsilon }{1+\epsilon }%
\right) ^{\mu _{S}-1}}  \label{exact}
\end{equation}%
These results are exact for $t_{0}=0$, but, as we discussed, in many real
applications the distribution of $C$ is necessarily determined from a moving
time window. In order to understand how these results can be useful in the
latter case, we make some intuitive observations, followed by additional
theory. We already noted that, when $\mu <2$, the mean length of the laminar
regions of a system is infinity. This explains why we obtain the erratic
plot in the left panel of Fig. (\ref{fig_twocubes}): for most of the
duration of the time series there are no events, thus the cross-correlation
is either 1 or -1. This fact can be exploited to obtain the regular behavior
of the right panel of Fig. (\ref{fig_twocubes}): when one of the two systems
has an event, it is most probably embedded in a long laminar region of the
other system. If the P-system has an event, then it is most likely embedded
in a long laminar region of the S-system. In this case the resulting value
of $C$ follows the statistics of the unperturbed Lamperti distribution given
by Eq. (\ref{ML2}), as the S-system has no influence on the P-system and the
latter is non-ergodic. If the S-system has an event, then it is most likely
embedded in a long laminar region of the P-system, which is equivalent to
saying that S is subject to constant stimulation. In this case the
computed value of $C$ follows the statistics of the perturbed Lamperti
distribution given by Eq. (\ref{U}). The probability $W_{S}$ of having an
event in S at time $t$ is given by: 
\begin{equation}
W_{S}(t)=\frac{R_{S}(t)}{R_{S}(t)+R_{P}(t)},  \label{eqProbS}
\end{equation}
with $R(t)$ given by Eq. (\ref{fellercascade}) with the $\mu $ of the
corresponding system. The probability $W_{P}$ can be obtained from (\ref{eqProbS}) by exchanging
the roles of S and P. When $t\rightarrow \infty $, if $\mu _{S}>\mu _{P}$,
we have $W_{S}=1$ and $W_{P}=0$; if $\mu _{S}<\mu _{P}$ then $W_{S}=0$ and $%
W_{P}=1$. As a side note we observe that this argument implies that the
perturbed system does not respond asymptotically to simple perturbations,
which corresponds to the phenomenon of habituation. 

The red stripes superimposed on the numerical calculations in the right
panel of Fig (\ref{fig_twocubes}) are determined using Eqs. (\ref{exact}) and (\ref{eqProbS})
and show excellent agreement with the numerical simulations. The above derivation is valid also in the case in
which one of the systems is ergodic and the other is not ergodic: in the
long time limit, only the former has events. This fact and the
considerations above imply that, in complete agreement with PCM, the
response of an ergodic system to a non-ergodic system is maximal. On the
other hand, the response of a non-ergodic system to an ergodic system
vanishes. In the case in which both sytems are ergodic, the above theory is
not applicable, but, given the equivalence (by definition) of ensemble
averages and time averages, in this case we again recover the results of
PCM, as expected.

In conclusion, this Letter extends GLRT, by indicating how to apply PCM to
single time series and determining how information is transfered between
systems. The issue was addressed at a formal level, in order to provide
results that are valid for a range of systems, that is, systems in the
physical, social and life sciences. These guidelines can be used to apply
the PCM to real experimental data, so as to assess, for instance, the
response of the brain to noninvasive stimuli, with the condition of
analyzing the crucial renewal events of the brain that are detectable and
observable, as shown by \cite{allegrini09,allegrini07}. In the literature
there is an increasing interest in criticality as well as in
intelligence-induced criticality \cite%
{chialvo10,haimovici13,beggs15,couzin07,plenz14} and the theory along with
the practical rules to detect correlation in the non-ergodic case, afforded
by this Letter, are expected to contribute to the advance of this field of
research.

\emph{acknowledgment} NP, DL and GP thank Welch for financial support
through Grant No. B-1577 and ARO for financial support through Grant
W911NF-15-1-0245.

\end{document}